\documentclass[fleqn]{aa}
\usepackage{amsmath}
\usepackage{amssymb,amsfonts}
\usepackage{eucal}
\usepackage{eufrak}
\usepackage{astron}
\usepackage{epsfig}
\newcommand*{\figOCS}{
  \begin{figure}[h]
    \begin{center}
      \epsfig{file=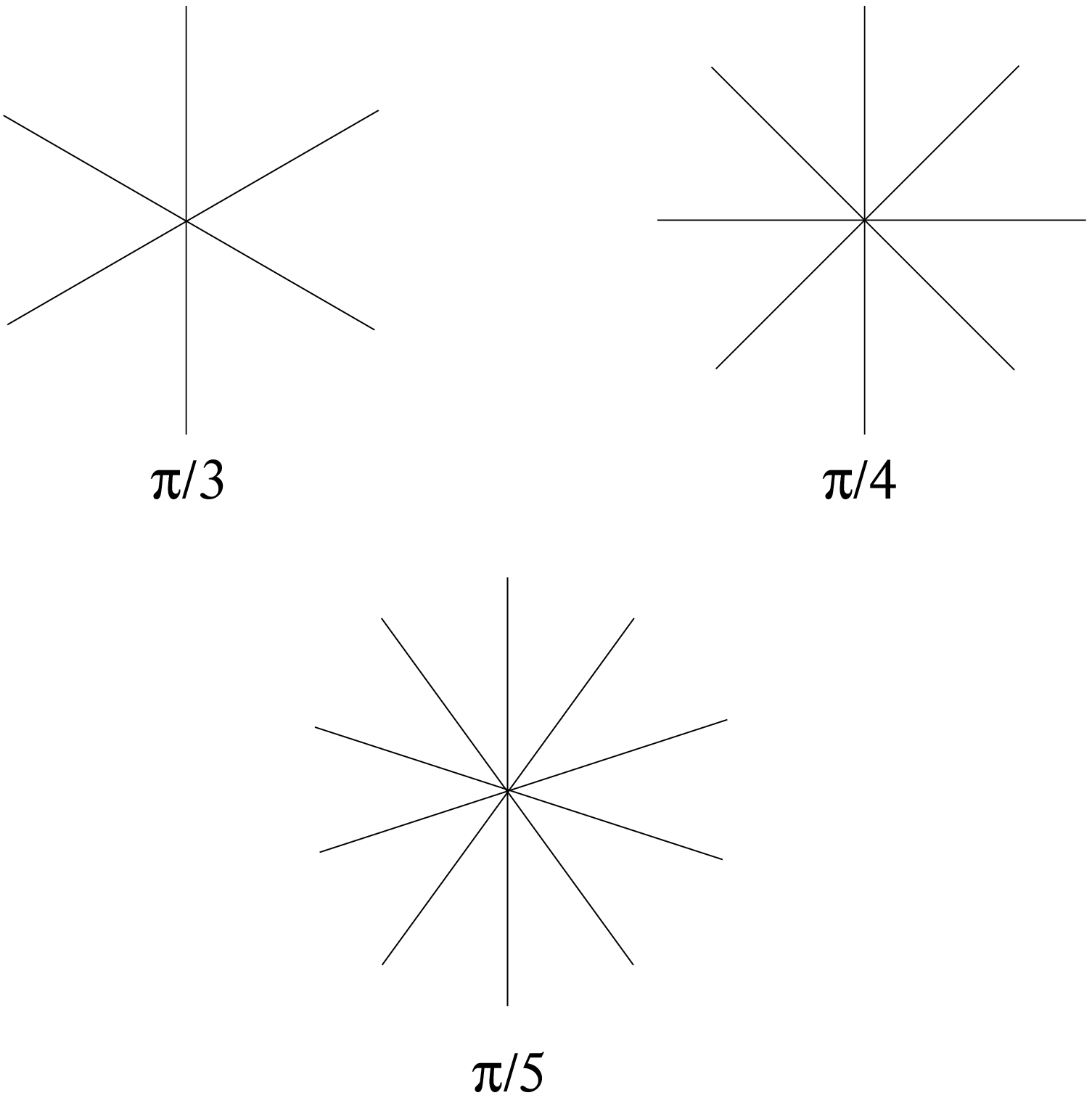, width=.45\textwidth}
    \end{center}
    \caption{The relative orientations of polarimeters in \OCs\ with
      3, 4 and 5 detectors. 
      \label{figOCS}}
  \end{figure}}
\newcommand*{\ns}{\!\!\!\!\!\!\!\!\!}
\newcommand*{\nss}{\!\!\!\!\!\!}
\newcommand*{\bc}{\begin{center}}
\newcommand*{\ec}{\end{center}}
\newcommand*{\OCs}{``Optimised Configurations''}

\newcommand*{\id}{\bbbone}

\newcommand*{\AB}{\boldsymbol{A}}

\newcommand*{\bb}{\boldsymbol{\beta}}

\newcommand*{\gb}{\boldsymbol{\gamma}}
\newcommand*{\MB}{\boldsymbol{M}}
\newcommand*{\NB}{\boldsymbol{N}}
\newcommand*{\nv}{\boldsymbol{\hat{n}}}

\newcommand*{\RB}{\boldsymbol{R}}
\newcommand*{\SB}{\boldsymbol{S}}
\newcommand*{\VB}{\boldsymbol{V}}
\newcommand*{\XB}{\boldsymbol{X}}

\newcommand*{\BC}{\boldsymbol{\mathcal{B}}}
\newcommand*{\GC}{\boldsymbol{\mathcal{G}}}

\begin{document}
\thesaurus{12(12.03.1;12.03.3;03.09.5;03.13.5;02.16.2) }
\bibliographystyle{astron}
\title{Optimised polarimeter configurations for measuring the Stokes
  parameters of the Cosmic Microwave Background Radiation}
\author{F.~Couchot\inst{1} \and
  J. Delabrouille\inst{2} \and  J. Kaplan\inst{3} \and B. Revenu\inst{3}
}
\institute{
Laboratoire de l'Acc{\'e}l{\'e}rateur Lin{\'e}aire, IN2P3 CNRS, Universit{\'e} 
Paris Sud, 91405 Orsay, France
\and
Institut d'Astrophysique Spatiale, CNRS \& Universit{\'e} Paris XI, 
b{\^a}t 121, 91405 Orsay Cedex, France, and \\
University of Chicago, Dept. of Astronomy and Astrophysics, 5640 South
Ellis Avenue, Chicago, IL 60637 USA 
\and
Physique Corpusculaire et Cosmologie, Coll{\`e}ge de France, 11 Place Marcelin Berthelot, 75231 Paris Cedex 05,  
France
}
\offprints{J. Kaplan (kaplan@cdf.in2p3.fr)}


\authorrunning{Couchot et al.}
\titlerunning{Optimised configurations of polarimeters for
  Measuring the CMB Stokes parameters}
\maketitle
\begin{abstract}
We present configurations of polarimeters which measure the three Stokes
parameters $I$, $Q$ and $U$ of the Cosmic Microwave Background
Radiation with a nearly 
diagonal error matrix, independent of the global orientation of the
polarimeters in the focal plane. These configurations also provide the
smallest possible error box volume. 
\end{abstract}
\section{Introduction} 
This paper originates from preparatory studies for the Planck satellite
mission. This Cosmic Microwave Background (CMB) mapping satellite is
designed to be able to measure the polarisation of the CMB in several frequency
channels with the sensitivity needed to extract the expected
cosmological signal. 
Several authors (see for instance
Rees\cite*{rees68a}, Bond and Efstathiou\cite*{bond87}, Melchiorri and
Vittorio\cite*{melchiorri96}, Hu and White\cite*{hu97b}, Seljak and Zaldarriaga\cite*{seljak98}),
have pointed out that   measurements of the polarisation of the CMB
will help to discriminate between cosmological models and
to separate the foregrounds. In the theoretical analyses of the polarised power
spectra, it is in general assumed (explicitly or implicitly) that the errors
are uncorrelated between the three Stokes parameters $I$, $Q$
and $U$\footnote{For a definition of the Stokes parameters, see for
  instance Born and Wolf \cite*{born80} page 554, where $I$, $Q$, $U$ are
noted $s_0$, $s_1$, $s_2$ respectively} in the 
reference frame used to build the polarised multipoles
\cite{zaldarriaga97,ng97}. However, the errors in the three Stokes 
parameters will in general be correlated, even if the noise of the
three or more measuring polarimeters are not, unless the layout of the
polarimeters is adequately chosen.  In this paper we construct
configurations of the relative orientations of the polarimeters,
hereafter called \OCs\ (OC), such that, if the noise in all polarised
bolometers have the same variance and are not correlated, the
measurement errors in the 
Stokes parameters $I$, $Q$ and $U$ are independent of the direction of
the focal plane and decorrelated. Moreover, the volume of the error
box is minimised. The properties of decorrelation and minimum
error are maintained when one combines redundant measurements of the same point of the
sky, even when the orientation of the focal plane is changed between
successive measurements. Finally, when combining unpolarised and data
from OC's, the resulting errors retain their optimised properties.

In general, the various polarimeters will not have the same levels of noise
and will be slightly cross-correlated. Assuming
that these imbalances and cross-correlations are small, we show that
for OC's\  the
resulting correlations between the errors on $I$, $Q$ and $U$ are also
small and easily calculated to first order. This remains
true when one combines several measurements on the same point of the
sky, the correlations get averaged but
do not cumulate.

Finally, we calculate the error matrix between $E$ and $B$ multipolar
amplitudes and show that it is also simpler in OC's

\subsection{General considerations} In the reference frame where 
the Stokes parameters  $I$, $Q$, and $U$ are defined, the
intensity detected by a polarimeter rotated by an angle $\alpha$ with respect
to the $x$ axis is:  
\begin{equation}
\ns I_\alpha = \frac{1}{2} (I + Q\, \cos 2\alpha + U\, \sin
2\alpha).
\end{equation}
Because polarimeters only measure intensities, angle $\alpha$ can
be kept between 0 and $\pi$. To be able to separate the 3 Stokes
parameters, at least 3 polarised detectors are needed (or 1
unpolarised and 2 polarised), with angular
separations different from multiples of $\pi/2$. If one uses $n \ge 3$ polarimeters with orientations $\alpha_p,\ 1 \le p \le n$ for a given line
of sight, the Stokes parameters will be estimated by minimising the
$\chi^2$: 
\begin{equation}
\ns\chi^2 = \left(\MB - \AB \SB\right)^T\, \NB^{-1}\,\left(\MB - \AB
  \SB\right)  
\end{equation}
where $ \MB = 
\left(\begin{array}{c}
m_1\\ \vdots \\m_p\\ \vdots \\m_n 
\end{array}\right)$
 is the vector of measurements, and $\NB$ is their $n\times n$
 noise autocorrelation matrix. The $n\times 3$ matrix  
\begin{equation}
\ns\AB =\frac{1}{2} \left(\begin{array}{ccc}
1&\cos 2 \alpha_1&\sin 2 \alpha_1\\
 \vdots & \vdots & \vdots\\
1&\cos 2 \alpha_p&\sin 2 \alpha_p\\
 \vdots & \vdots & \vdots\\
1&\cos 2 \alpha_n&\sin 2 \alpha_n\\
\end{array}\right)\label{matA}
\end{equation}
 relates the
  results of the $n$ measurements to the vector of the Stokes
  parameters $\SB=\left(\begin{array}{c} I\\Q\\U \end{array}\right)$
  in a given reference frame, for instance a reference 
  frame fixed with respect to the focal instrument.
 If one looks in the same direction of the
sky, but with the instrument rotated by an angle $\psi$ in the focal
plane, the matrix $A$ 
is simply transformed with a rotation matrix of angle $2\, \psi$:
\begin{equation}
\ns\AB \rightarrow \AB\ \RB(\psi), \mbox{ with } \RB(\psi) = 
\left(\begin{array}{ccc}
1&0&0\\
0&\cos 2\psi&\sin 2\psi\\
0&-\sin 2\psi&\cos 2\psi 
\end{array}\right). \label{arot}
\end{equation} 
As is well known, the resulting estimation for the Stokes parameters
and their variance matrix $\VB$ are:
\begin{equation}
\begin{split}
\ns & \SB = \left(\AB^T\, \NB^{-1}\, \AB\right)^{-1}\, \AB^T \NB^{-1}
\MB, \\
\ns & \mbox{and } \\
\ns & \VB = \left(\AB^T\, \NB^{-1}\, \AB\right)^{-1} 
\end{split}
\label{variance}
\end{equation}
\section{Optimised Configurations}
\subsection{The ideal case}
If we assume that the measurements $m_p\ (1 \le p \le n)$ have identical
 and decorrelated errors \\($\NB_{p q} = <\delta m_p\ \delta
 m_q> = \sigma_0^2\, 
 \delta_{p q}$), the $\chi^2$ is simply: 
 \begin{equation}
\ns\chi^2 = \frac{1}{\sigma^2} \sum_{p=1}^n\left[m_p -
  \frac{1}{2}(I + Q\,\cos 2\alpha_p + U\,\sin 2\alpha_p)\right]^2,
\end{equation}
and the inverse of the covariance matrix of the Stokes parameters is  given by:
\begin{gather}
\ns \VB^{-1}=\frac{1}{\sigma^2}\,\XB, \qquad\qquad \XB = \AB^T\AB
= \frac{1}{4} \ \times \label{covar} \\
\ns \left(\begin{array}{lll}
n                                 & \sum_{1}^n\,\cos 2\alpha_p
&  
\sum_{1}^n\,\sin 2\alpha_p \\
&& \\
\sum_{1}^n\,\cos 2\alpha_p  & \frac{1}{2}(n + \sum_{1}^n\,\cos 4\alpha_p) &
\frac{1}{2}\sum_{1}^n\,\sin 4\alpha_p \\
&&\\
\sum_{1}^n\,\sin 2\alpha_p  & \frac{1}{2}\sum_{1}^n\,\sin 4\alpha_p       &
\frac{1}{2}(n - \sum_{1}^n\,\cos 4\alpha_p) 
\end{array}\right). \notag
\end{gather}
It is shown in the appendix that, if the orientations of the
polarimeters are evenly distributed on 180$^\circ$:
\begin{equation}
\ns\alpha_p = \alpha_1 + (p-1)\,\frac{\pi}{n},\ p = 1...n,\mbox{ with } n
\ge 3, \label{DC1}  
\end{equation}
the matrix $\VB$ takes the very simple diagonal form:
\begin{equation}
\label{vzero}
\ns\VB_0 = \sigma^2 {\XB_0}^{-1}, \mbox{ with } {\XB_0}^{-1} = \frac{4}{n}
\left(\begin{array}{ccc}
1 & 0 & 0 \\
0 & 2 & 0 \\
0 & 0 & 2 \\
\end{array}\right), 
\end{equation}
independent of the orientation of the focal plane, while its determinant is
minimised.
In other words, at the same time, the errors on the Stokes parameters
get decorrelated, their error matrix becomes independent of the
orientation of the focal plane and the volume of the error ellipsoid takes its
smallest possible value : $\frac{\pi}{3} \left(\frac{4\sigma}{\sqrt{n}}\right)^3$. 

The \OCs\ (OC) are the sets of polarimeters
which satisfy condition 
(\ref{DC1}),  (see figure \ref{figOCS}).\figOCS  They are hereafter referred to by the subscript $0$ as in equation
\eqref{vzero}. 
The smallest OC involves three polarimeters with relative angle
$\pi/3$. 
With 4 polarimeters, the angular separation must be $\pi/4$, and so
on. Note that 
a configuration with one unpolarised detector  
and 2 polarised detectors can never measure the Stokes parameters with
uncorrelated errors, because this would require polarimeters oriented 
90$^\circ$ apart from each other, which would not allow the breaking of the
degeneracy between $Q$ and $U$.  

If one combines several OC's\ with several unpolarised
detectors, all uncorrelated with each other, the resulting covariance
matrix for the Stokes  
parameters remains diagonal and independent of the orientation of 
the various OC's. More precisely, when combining the measurements 
of $n_T$ unpolarised detectors (temperature measurements),  with $n_P$ 
polarised detectors arranged in OC's,
the covariance matrix of the Stokes parameters reads:
\begin{equation}
\ns\VB\, =\, \frac{4\,\sigma_P^2}{n_P}
\left(\begin{array}{ccc}
\left(1 +
  4\,\frac{n_T}{n_P}\left(\frac{\sigma_P}{\sigma_T}\right)^2\right)^{-1}
& 0 & 0 \\
0 & ~~~2~~~ & 0 \\
0 & 0 & ~~~2~~~ \\
\end{array}
\right),\label{vzero1}
\end{equation}\\
where we have introduced inverse average noise levels,
$\sigma_T$  and $\sigma_P$, for the unpolarised and polarised
detectors respectively:  
\begin{equation}
\ns\frac{1}{{\sigma_T}^2} = \left\langle
  \frac{1}{\sigma_\mathrm{unpolarised}^2}\right\rangle, \mbox{ and  }
\frac{1}{{\sigma_P}^2} =
\left\langle\frac{1}{\sigma_\mathrm{polarised}^2}\right\rangle .  
\end{equation}
Note that the levels of noise can be different from one OC
to the other and from those of the unpolarised detectors.

\subsection{A more realistic description of the measurements}
In general one expects that there will be some slight imbalance and
cross-correlation between the noise of the detectors.
The noise matrix of the measurements will in general take the form:
\begin{equation}
\ns\NB = \sigma^2 (\id + \bb + \gb) \label{nideal},
\end{equation}
where the imbalance $\bb$ and cross-correlation $\gb$ matrices
\begin{equation}
\begin{split}
\ns & \bb = 
\left(\begin{array}{cccc}
\beta_{1\,1}&0            &0            &\ldots\\
0            &\beta_{2\,2}&0            &\ldots\\
0            &0            &\beta_{3\,3}&\ldots\\
\vdots       &\vdots       &\vdots       &\ddots\\
\end{array}\right),\  \mathrm{Tr}(\bb) = 0, \mbox{ and }\\
\ns & \gb= 
\left(\begin{array}{cccc}
0            &\gamma_{1\,2}&\gamma_{1\,3}&\ldots\\
\gamma_{1\,2}&0            &\gamma_{2\,3}&\ldots\\
\gamma_{1\,3}&\gamma_{2\,3}&0            &\ldots\\
\vdots       &\vdots       &\vdots       &\ddots\\
\end{array}\right),
\end{split}
\end{equation}
are ``small'', that is will be treated as first order perturbations
in the following, and therefore 
\[
\ns \NB^{-1} = \frac{1}{\sigma^2}\, (\id - \bb - \gb).
\]
To this order, the variance matrix of the Stokes parameters is easily
obtained from Eq. (\ref{variance}):
\begin{equation}
\ns\VB = \sigma^2 \left[\XB^{-1} + \BC + \GC \right],
\end{equation}
where the matrix $\XB$ is given by Eq. \eqref{covar} and the first
order corrections to $\VB$, matrices $\BC$ and $\GC$, read:
\begin{equation}
\ns\left(\begin{array}{l}
\BC\\
\GC
\end{array}\right)
 = \XB^{-1}\ \AB^T\,
\left(\begin{array}{l}
\bb\\
\gb
\end{array}\right)
\,\AB\ \XB^{-1}. \label{perturb1}
\end{equation}
In an OC, the matrix  $\XB^{-1}$ takes the simple diagonal form
${\XB_0}^{-1}$ of 
Eq. (\ref{vzero}), and the non diagonal parts, $\BC$ and $\GC$ remain
of order 1 in  $\bb$ and $\gb$. For instance, if we consider an OC
with 3  polarimeters, and polarimeter number 1 is oriented in the $x$
direction,
\[
\ns{\BC} = \frac{4}{3}\
\left(\begin{array}{ccc}
0               &  \beta_{1\,1} & \frac{\beta_{2\,2} - \beta_{3\,3}}{\sqrt{3}} \\
\beta_{1\,1}   &  \beta_{1\,1} & \frac{\beta_{3\,3} - \beta_{2\,2}}{\sqrt{3}} \\
\frac{\beta_{2\,2} - \beta_{3\,3}}{\sqrt{3}} &\frac{\beta_{3\,3} -
  \beta_{2\,2}}{\sqrt{3}} & - \beta_{1\,1}                                  \\
\end{array}\right),\, 
\]
 where $\beta_{2\,2} + \beta_{3\,3} = - \beta_{1\,1}$, and 
\[
\begin{split}
\ns  \GC = \frac{4}{3} &\times \\
\ns  & \left(\begin{array}{ccc}
\frac{2(\gamma_{1\,2} + \gamma_{1\,3} + \gamma_{2\,3})}{3} & \frac{\gamma_{1\,2} + \gamma_{1\,3} - 2 \gamma_{2\,3}}{3} & 
\frac{\gamma_{1\,2} - \gamma_{1\,3}}{\sqrt{3}} \\
\frac{\gamma_{1\,2} + \gamma_{1\,3} - 2 \gamma_{2\,3}}{3} & \frac{2(\gamma_{2\,3} - 2 \gamma_{1\,2} -2 \gamma_{1\,3})}{3} & 
\frac{2 (\gamma_{1\,2} - \gamma_{1\,3})}{\sqrt{3}} \\
\frac{\gamma_{1\,2} - \gamma_{1\,3}}{\sqrt{3}} &\frac{2 (\gamma_{1\,2}
  - \gamma_{1\,3})}{\sqrt{3}} & - 2\,\gamma_{2\,3}  
\end{array}\right).
\end{split}
\]
Note that $\BC$ and $\GC$ transform under a rotation of the focal
plane by a rotation $\RB(\psi)$:
\begin{equation}
\ns\left(\begin{array}{l}
\BC\\
\GC
\end{array}\right)
\rightarrow {\RB(\psi)}^{-1}
\left(\begin{array}{l}
\BC\\
\GC
\end{array}\right) \RB(\psi).
\end{equation}
Because $\VB_0$ is invariant, as long as $\bb$ and $\gb$ are
small the dependence of $\VB$ on the orientation of the focal plane
remains weak.

\section{Co-adding measurements}                               
The planned scanning strategy of Planck goes stepwise: at each
step the satellite will spin about 100 times around a fixed axis,
covering the same  circular scan, then the spin axis of the satellite
will be moved by a few arc-minutes, and so on. This
provides two types of redundancy: every pixel along each circle will
be scanned about 100 times, and some pixels will be seen by
several circles, with different orientations of the focal plane. In
this section we show, assuming a perfect white noise 
along each scan, that the properties of the error
matrix of the Stokes parameters  coming from OC's\  are kept if
all data are simply co-added at each pixel, whatever the
orientations of the focal plane. The redundancy provided by
intersecting circles can be used to remove the stripes induced on maps by  
low-frequency noise in the data streams. 
An extension adapted to polarised measurements of the method proposed by 
Delabrouille \cite*{delabrouille98a} for the 
de-striping of Planck maps is studied in Revenu et al. \cite*{revenu98}.

Here we assume that the noise is not correlated between different scans
and can thus be described by one matrix $\NB_l$ for each scan, indexed
by $l$,  passing 
through the pixel. The $\chi^2$ is then the sum of the  $\chi^2_l$
over the $L$ scans that cross the pixel: 
\begin{equation}
\ns\chi^2 = \sum_{l=1}^L (\MB_l - \AB_l\, \SB)^T {\NB_l}^{-1} (\MB_l - \AB_l\, \SB).
\label{chi2l}
\end{equation}
 The estimator of the Stokes parameters stemming from this $\chi^2$ is
\begin{equation}
\ns\SB = \left(\sum_{l=1}^L \AB_l^T\,
  \NB_l^{-1}\,\AB_l\right)^{-1}\,\sum_{l=1}^L\AB_l^T\,
\NB_l^{-1}\, \MB_l, 
\end{equation} 
with  variance matrix:
\begin{equation}
\ns\VB = \left(\sum_{l=1}^L \AB_l^T\, \NB_l^{-1}\,\AB_l\right)^{-1}.
\end{equation}
\paragraph{In the ideal case,} for a given scan, the noise 
(assumed to be white on each scan) has the same variance for all bolometers
with no correlation between them,
although it can vary from one scan to the other:
\begin{equation}
\ns\NB_l = {\sigma_{l}}^2 \id, 
\end{equation}
and one can write the resulting variance combining the $L$ scans:
\begin{equation}
\ns\VB_L = \left(\sum_{l=1}^L \frac{\XB_l}{{\sigma_l}^2}\right)^{-1} = 
\left(\sum_{l=1}^L \frac{1}{{\sigma_l}^2}
  \RB^{-1}(\psi_l)\,\XB_1\, \RB(\psi_l)\right)^{-1}\label{vlideal}, 
\end{equation}
where $\XB_l = \AB_l^T\,\AB_l$, and we have written explicitly the rotation
matrices which connect the orientation of the focal plane along scan $l$
with that along scan $1$. Note that these matrices are dependent of
the position along the scan through angle $\psi_l$ .

If the observing setup is in an
OC, all orientation dependence drops out and the expression of
the covariance matrix becomes diagonal as for a single measurement
(Eq. \ref{vzero}):
\begin{equation}
\ns \VB_{0\,L}  =  
\frac{4\,{\sigma_L}^2}{n\,L}
\left(\begin{array}{ccc}
1 & 0 & 0 \\
0 & 2 & 0 \\
0 & 0 & 2 \\
\end{array}
\right) = \frac{{\sigma_L}^2}{L}\,{\XB_0}^{-1}, \label{vlzero}
\end{equation}
where ${\XB_0}^{-1}$ is defined in Eq. \eqref{vzero} and the average noise
level $\sigma_L$ is defined as: 
\begin{equation}
\ns\frac{1}{{\sigma_L}^2} = \left\langle\frac{1}{{\sigma_l}^2}\right\rangle. 
\end{equation}
Of course one recovers the fact that, with $L$ measurements, the errors on
the Stokes parameters are reduced by a factor~$\sqrt{L}$.

\paragraph{More realistically,} we expect that the noise matrices
will take a form similar to Eq.~\eqref{nideal}:
\begin{equation}
\ns\NB_l = {\sigma_l}^2 \left(\id + \bb_l +
  \gb_l\right) \label{vlnideal}  
\end{equation}
If $\bb_l$ and $\gb_l$ are small, first order inversion allows to
calculate $\VB$ ($\VB_L$ is given by Eq. \eqref{vlideal}): 
\begin{equation}
\ns\VB = \VB_L +\VB_L \sum_{l=1}^L \AB_l^T  \frac{\bb_l +
  \gb_l}{{\sigma_l}^2} \AB_l\VB_L.
\end{equation}
If the focal plane is in an OC, this expression simplifies to
\begin{equation}    
\ns\VB_{L} =
\frac{{\sigma_L}^2}{L}\,
\left({\XB_0}^{-1} +
  \frac{{\sigma_L}^2}{L}\,\sum_{l=1}^L \frac{\BC_l+\GC_l}{{\sigma_l}^2}\right),
\label{vlpert}
\end{equation}
 where
\begin{equation}
\ns \left(\begin{array}{l}
\BC_l\\
\GC_l
\end{array}\right)
 = \RB^{-1}(\psi_l) {\XB_0}^{-1}\ \AB_1^T\,
\left(\begin{array}{l}
\bb_l\\
\gb_l
\end{array}\right)
\,\AB_1\ {\XB_0}^{-1} \RB(\psi_l).
\end{equation}

The $1/L$ factor
inside the parenthesis in equation (\ref{vlpert}) implies that the
cross-correlations and the dependence on  on the orientation $\psi_l$
of the focal plan remain weak when one cumulates measurements of the
same pixel. 

\section{The error covariance matrix of the scalar $E$ and $B$ parameters}

Scalar polarisation parameters, denoted $E$ and $B$, have been
introduced, which do not
depend on the reference frame \cite{newman66,zaldarriaga97}. However,
the properties 
of OC's do not propagate simply to the
error matrix of the $E$ and $B$ parameters  because 
their definition is non local in terms of the Stokes
parameters.

Nevertheless, if the measurements errors are not correlated
between different points of the sky (or if the correlation has
been efficiently suppressed by the data treatment) then the elements of the
error matrix 
of the multipolar coefficients $a_{E,lm}$ and
$a_{B,lm}$ 
 can be written in a simple form which is given in Appendix B for a general configuration.

For an OC, the error matrix
simplifies further and its elements reduce to:
\[
\begin{split}
\ns & <\delta a_{^E_B, lm}\,{\delta a_{^E_B, l'm'}}^*> = \frac{1}{2} \left(\frac{4\,\pi}{N_{pix}}\right)^2
\sum_{\nv_k} \sigma_{\mathrm{Stokes}}^2(\nv_k)\\
\ns & ~~~~~\times [{{}_2Y_{lm}(\nv_k)}^*\,{}_2Y_{l'm'}(\nv_k) +
{{}_{-2}Y_{lm}(\nv_k)}^*\,{}_{-2}Y_{l'm'}(\nv_k)]\\
\ns & <\delta a_{E, lm}\,{\delta a_{B, l'm'}}^*> = \frac{i}{2} \left(\frac{4\,\pi}{N_{pix}}\right)^2
\sum_{\nv_k}\sigma_{\mathrm{Stokes}}^2(\nv_k) \\
\ns & ~~~~~\times [{{}_2Y_{lm}(\nv_k)}^*\,{}_2Y_{l'm'}(\nv_k) -
{{}_{-2}Y_{lm}(\nv_k)}^*\,{}_{-2}Y_{l'm'}(\nv_k)].
\end{split}
\]
where $N_{pix}$ is the total number of pixels in the sky, $\sigma^{}_{\mathrm{Stokes}}$ is the common r.m.s. error on the
$Q$ and $U$ Stokes parameters,  $\nv_k$ is the direction of pixel
$k$ and functions 
$_{\pm 2}Y_{lm}(\nv_k)$ are the spin 2 spherical harmonics. If 
$\sigma^{}_{\mathrm{Stokes}}$  does 
not depend on the direction in the sky, a highly improbable situation, 
then the
orthonormality of the spin weighted spherical harmonics makes the error
matrix fully diagonal in the limit of a large number of pixels:
\[\begin{split}
\ns & \left( \begin{array}{cc}
<\delta a_{E, lm}\,{\delta a_{E, l'm'}}^*> & <\delta a_{E, lm}\,{\delta a_{B, l'm'}}^*> \\
<\delta a_{B, lm}\,{\delta a_{E, l'm'}}^*> & <\delta a_{B, lm}\,{\delta a_{B, l'm'}}^*>
\end{array} \right) \\
\ns & ~~~~~~~~~~~~~~~~~~~~~~~~~~~~~~~~~~~~~~ = \id\ \frac{4\,\pi}{N_{pix}}\ \sigma_{\mathrm{Stokes}}^2\ \delta_{ll'}
\delta_{mm'}. 
\end{split}\]
Note that, even for unpolarised data, the error matrix between multipolar
amplitude is not diagonal unless the r.m.s. error is constant over the
whole sky (see for instance  Oh, Spergel and Hinshaw, 1998). 

In the same conditions, the noise matrix of fields $E$ and $B$ is also
 fully diagonal:
\[
\ns \left(\begin{array}{cc}
\left<\delta E(\nv)\ \delta E(\nv')\right> & \left<\delta E(\nv)\ \delta B(\nv')\right>\\
\left<\delta B(\nv)\ \delta E(\nv')\right> & \left<\delta B(\nv)\ \delta B(\nv')\right>
\end{array}\right) = \id\ \sigma_{\mathrm{Stokes}}^2\ \delta_{\nv\, \nv'}.
\]

\section{Conclusion}
In this paper we have shown that, if the noise of the polarimeters
have nearly equal levels and are approximately uncorrelated, these detectors
can be set up in ``Optimised Configurations''. These
configurations are optimised in two respects: first the volume of the
error ellipsoid is minimised, and second the 
error matrix of the inferred Stokes parameters is approximately
diagonal in all reference frames. This remains true even if one
combines information from numerous measurements along different
scanning circles. Such \OCs, with 3 and 4 polarimeters, have
been proposed by the Planck High Frequency Instrument consortium
\cite{planckao97} for the three channels where it is intended to
measure the CMB polarisation. 

\appendix
\section{The conditions for an OC} \label{ApA}
In this appendix, we show that conditions \eqref{DC1} diagonalise
the error matrix $\VB$ of the Stokes parameters and minimise
its determinant, if the noises in the $n$ polarimeters have identical
levels and are not correlated. \\
We use the notation:
\[
\ns S_k = \sum_{p=1}^{n}\,e^{i\,k\,\alpha_p} = |S_k|\,e^{i\,\theta_k}, \ k = 2,4.
\]
It can be seen from equation \eqref{covar} that requiring that the
error on $I$ be decorrelated from the errors on $Q$ and $U$ is
equivalent to the condition: 
\begin{equation}
\ns S_2 = 0.\label{cond0}
\end{equation} 
This condition can easily be fulfilled in a configuration where the
angles $\alpha_p$ are regularly distributed:
\begin{equation}
\begin{split}
\ns & \alpha_p = \alpha_1 + (p-1)\,\delta\alpha,\ p = 1...n,\\
\ns & \mbox{with } \ n \ge 3,\ 0 <
\delta\alpha < \pi, \ \delta\alpha \ne \pi/2\ \mbox{ (see text)}.
\end{split}
\label{ADC1}
\end{equation}
In such configurations, equation (\ref{cond0}) becomes: 
\begin{equation}
\ns S_2 =
e^{i\,2\,\alpha_1}\frac{e^{i\,2\,n\,\delta\alpha}-1}{e^{i\,2\,\delta\alpha}-1}
= 0,
\label{cond1} 
\end{equation}
The solutions of  equation (\ref{cond1}) under conditions (\ref{ADC1})
reduce to: 
\begin{equation}
\ns\delta\alpha = \frac{\pi}{n}, \mbox{ with } n \ge 3. \label{ADC2}
\end{equation}
It is easily seen that conditions (\ref{ADC2}) also automatically
ensure that $S_4 = 0$
and therefore that $\XB$ becomes diagonal and assumes the very
simple form:  
\begin{equation}
\label{vzerom1}
\ns\XB_0 = \frac{n}{4}
\left(\begin{array}{ccc}
1 & 0 & 0 \\
0 & 1/2 & 0 \\
0 & 0 & 1/2 \\
\end{array}\right)
\end{equation}
independent of the orientation of the focal plane. Equation \eqref{vzero}
is the consequence of \eqref{vzerom1}

The error volume is proportional to the determinant of the error
matrix $\VB$. Therefore, it is minimum when the
determinant of $\XB$  (equation \eqref{covar}) is
maximum. This determinant can be written as:
\begin{multline}
\nss\mathrm{Det}( \XB) =\\
 \frac{1}{64}
\left[n^3 - n\,|S_4|^2 - 2\,  |S_2|^2(n - |S_4|\, \cos(\theta_4 - 2\,
  \theta_2))\right].\label{A1} 
\end{multline}
Because the $S_k$'s are sums of $n$ complex numbers with modulus 1, $|S_k|
< n$, and it is clear from Eq.~\eqref{A1} that 
\[
\mathrm{Det}(\XB) \le \frac{n^3}{64},
\]
and that the upper bound is reached if and only if
\begin{equation}
S_2 = S_4 = 0. \label{A2}
\end{equation}
Conditions \eqref{ADC1} and \eqref{ADC2} have been shown above to
imply \eqref{A2}, and  therefore ensure that 
the determinant of the covariance matrix $\VB$ is minimum.

\section{The general error matrix of the $E$ and $B$ multipolar
  coefficients}
\label{ApB}
Assuming that the data treatment has removed all correlation between
different directions in the sky, the matrix elements of the error matrix 
of the  $E$ and $B$ multipolar coefficients are:
\begin{equation}
\begin{split}
\ns & <\delta a_{^E_B, lm}\,{\delta a_{^E_B, l'm'}}^*> = \frac{1}{4} \left(\frac{4\,\pi}{N_{pix}}\right)^2
\sum_{\nv_k}\\
\ns & (\NB_{QQ} + \NB_{UU})(\nv_k) [{{}_2Y_{lm}}^*\,{}_2Y_{l'm'} +
{{}_{-2}Y_{lm}}^*\,{}_{-2}Y_{l'm'}](\nv_k)\\
\ns & \pm (\NB_{QQ} - \NB_{UU})(\nv_k) [{{}_2Y_{lm}}^*\,{}_{-2}Y_{l'm'} +
{{}_{-2}Y_{lm}}^*\,{}_{2}Y_{l'm'}](\nv_k)\\
\ns & \pm 2i\,\NB_{QU}(\nv_k) [{{}_2Y_{lm}}^*\,{}_{-2}Y_{l'm'} -
{{}_{-2}Y_{lm}}^*\,{}_{2}Y_{l'm'}](\nv_k),\\
\ns & <\delta a_{E, lm}\,{\delta a_{B, l'm'}}^*> = \frac{i}{4} \left(\frac{4\,\pi}{N_{pix}}\right)^2
\sum_{\nv_k}\\
\ns & (\NB_{QQ} + \NB_{UU})(\nv_k) [{{}_2Y_{lm}}^*\,{}_2Y_{l'm'} -
{{}_{-2}Y_{lm}}^*\,{}_{-2}Y_{l'm'}](\nv_k)\\
\ns & - (\NB_{QQ} - \NB_{UU})(\nv_k) [{{}_2Y_{lm}}^*\,{}_{-2}Y_{l'm'} -
{{}_{-2}Y_{lm}}^*\,{}_{2}Y_{l'm'}](\nv_k)\\
\ns & - 2i\,\NB_{QU}(\nv_k) [{{}_2Y_{lm}}^*\,{}_{-2}Y_{l'm'} +
{{}_{-2}Y_{lm}}^*\,{}_{2}Y_{l'm'}](\nv_k), \nonumber
\end{split}
\end{equation}
where $\NB(\nv_k)$ is the noise matrix of the Stokes
parameters $Q$ and $U$ in the direction $\nv_k$ of pixel~
$k$. 

\begin{acknowledgement}
We thank Alex Kim for useful suggestions and for
reading the manuscript, and the referee, Uro\v s Seljak, who suggested
to consider the implications on the $E$ and $B$ scalar parameters.
\end{acknowledgement}

\end{document}